\documentclass[reprint, amsmath, amssymb, aps, prfluids, superscriptaddress]{revtex4-2}

\usepackage{graphicx}
\usepackage{bm}
\usepackage{booktabs}
\usepackage{hyperref}
\usepackage{xcolor}
\usepackage{multirow}%
\usepackage{amsmath,amssymb,amsfonts}%
\usepackage{mathrsfs}%
\usepackage{textcomp}%

\newcommand{\ndP}{\widetilde{P}}

\newcommand{\ndt}{\tilde{t}}

\newcommand{\fvel}{v}
\newcommand{\ndL}{\tilde{L}}

\newcommand{\ee}{\mathrm{e}}

\newcommand{\vx}{\mathbf{x}}
\newcommand{\vxz}{\mathbf{x}_0}

\newcommand{\vvr}{\mathbf{r}}
\newcommand{\vv}{\mathbf{v}}
\newcommand{\ve}{\mathbf{e}}
\newcommand{\vd}{\mathbf{d}}
\newcommand{\vF}{\mathbf{F}}
\newcommand{\vf}{\mathbf{f}}
\newcommand{\vk}{\mathbf{k}}

\newcommand{\vI}{\mathbf{I}}
\newcommand{\gradzero}{\nabla_0}

\begin{document}

\title{Mixing induced by microswimmers as probed by mutual information}

\author{Yihong Shi}
\email{yihong.shi@ista.ac.at}
\affiliation{Max Planck Institute for Dynamics and Self-Organization (MPI-DS), Am Fassberg 17, 37077 G\"{o}ttingen, Germany}
\affiliation{Institute of Science and Technology Austria, 3400 Klosterneuburg, Austria}

\author{Yuto Hosaka}
\affiliation{Max Planck Institute for Dynamics and Self-Organization (MPI-DS), Am Fassberg 17, 37077 G\"{o}ttingen, Germany}
\affiliation{Department of Mathematics, Kyoto University, 606-8502 Kyoto, Japan}

\author{Andrej Vilfan}
\affiliation{Max Planck Institute for Dynamics and Self-Organization (MPI-DS), Am Fassberg 17, 37077 G\"{o}ttingen, Germany}
\affiliation{Jo\v{z}ef Stefan Institute, 1000 Ljubljana, Slovenia}

\author{Ramin Golestanian}
\email{ramin.golestanian@ds.mpg.de}
\affiliation{Max Planck Institute for Dynamics and Self-Organization (MPI-DS), Am Fassberg 17, 37077 G\"{o}ttingen, Germany}
\affiliation{Rudolf Peierls Centre for Theoretical Physics, University of Oxford, Oxford OX1 3PU, United Kingdom}

\date{\today}

\begin{abstract}
We investigate fluid mixing induced by microswimmers using mutual information as a global, information-theoretic measure of mixing efficiency. For a two-dimensional squirmer model in a confined domain, we compute numerically the swimmer-generated flows and solve the advection–diffusion equation for the transport of tracer particles in the fluid. We show that the spatial distribution of swimmers strongly affects mixing, which is suppressed by swimmer aggregation and enhanced by positional and orientational disorder. At fixed energy dissipation, mixing efficiency depends non-monotonically on the squirmer parameter, with an optimal finite value arising from the balance between swimmer translation and dipolar flow generation. When hydrodynamic interactions are included, pushers outperform pullers. The mutual information as a function of time decays in three stages: an initial diffusion-dominated stage, an intermediate advection-enhanced regime, and a final relaxation stage controlled by system size.  Our results demonstrate that mutual information, previously validated as a measure of mixing efficiency only in simplified model systems, can equally be used in complex flows. Its application reveals that mixing by microswimmers is subject to a trade-off between the generation of strong shear flows and achieving optimal dispersion across the fluid domain.
\end{abstract}

\maketitle

\section{Introduction}\label{sec1}
Mixing describes the homogenization of the nonequilibrium scalar field (concentration or temperature) in a deforming continuum medium. Mixing fundamentally relies on molecular diffusion. However, because diffusion at large scales is intrinsically slow, efficient mixing requires additional advective stirring. In microscale environments, stirring alone cannot mix fluids because of the time-reversibility of flows at low Reynolds numbers \cite{Heller1960,Arrieta.Tuval2020,Golestanian2011}. Therefore, the interplay between advection (stirring) and diffusion is of paramount importance in mixing processes and needs to be taken into account in norms that quantify the efficiency of mixing \cite{Tang.Golestanian2020,Villermaux2019,8}. Conventional measures that quantify the mixing efficiency include global metrics such as the $L^2$ norm \cite{Danckwerts1952,Thiffeault2012} and Sobolev norms \cite{Mathew.Petzold2005,Thiffeault2012}, as well as local measures based on the amount of stretching or the Lyapunov exponents \cite{Arrieta.Tuval2020,Meunier.Villermaux2022,Tsang.Ott2005}.
%Here, we use mutual information as a global assumption-free measure for fluid mixing induced by microswimmers.
The information-theoretic measure {\em mutual information} was introduced in our previous work as a mathematically precise assumption-free measure of fluid mixing \cite{8}. Mutual information is universally applicable across a wide range of situations and does not depend on a specific pattern that is being erased by mixing. However, its practical application to mixing problems has so far been limited to simple model systems, such as Couette flow \cite{8} and uniform shear flow \cite{10}. 

Efficient fluid mixing at the microscale is essential in both biological and artificial systems, from the uptake of oxygen, nutrients and chemical signals in aquatic organisms \cite{Aref.Tuval2017,Shapiro.Stocker2014,Gilpin.Prakash2017} 
to microreactors and lab-on-a-chip technologies \cite{Aref.Tuval2017,Ottino.Grzybowski2004, Grigoriev.Sharma2006,Pine.Leshansky2005,Stroock.Whitesides2002}. At microscales, fluid mixing is often achieved through the beating of cilia, which can generate long-range flows as well as localized regions of chaotic advection \cite{Ding.Kanso2014,Jonas.Choma2013,Nawroth.McFallNgai2017,Shields.Superfine2010,Rahbar.Gray2014,Supatto.Vermot2008,Selvan.Brumley2026,Uchida2010}.
The activity of micro-organisms is also one of the primary factors enhancing fluid mixing at larger scales, which is then named biogenic mixing or biomixing \cite{Kunze2019}. 
Biogenic mixing refers to the enhancement of transport and homogenization of chemicals, heat or other organisms in fluids driven by the motion of living organisms. This process has been proposed as a potential mechanism for mixing in the ocean \cite{Huntley2004,Kunze2006, Katija2009,34,Katija2012,Wang2015,32} and investigated through both theoretical and experimental studies.
In previous studies, mixing induced by microswimmers has been primarily characterized through tracer particle displacements and associated effective diffusion coefficients \cite{Wu2000,33,30,Leptos2009,35,36, 31,Pushkin2014,Thiffeault2015,Mueller.Thiffeault2017,Ortlieb2019,Ishikawa2025}. Alternatively, a recent study of mixing by squirmers in stratified fluids defines mixing efficiency based on the energy transferred between swimmers and surrounding fluids \cite{32}. 

In this article, we apply mutual information between the initial and the final position of a particle in the fluid to quantify the mixing efficiency induced by microswimmers. We use the two-dimensional (2D) squirmer model to describe the flow induced by a non-deformable circular particle that swims in Stokes flow. This envelope model was originally introduced by Lighthill (1952) \cite{17} and further developed by Blake (1971) \cite{18}. %Within this framework, the swimmer type is characterized by the parameter $\beta$: a stresslet for $|\beta|=\infty$, a puller for $\beta>0$, a pusher for $\beta<0$, and a source dipole for $\beta=0$. Using the method of image,
We first solve the squirmer flow field confined to a 2D square domain and then compute the mutual information by solving the corresponding advection-diffusion equation. 
By examining the mixing efficiency of non-motile stresslets arranged in different spatial configurations, we show that swimmer aggregation suppresses fluid mixing, whereas disordered orientations and positions enhance it. Furthermore, we show that the optimal swimmer type that maximizes mixing efficiency at a given dissipation is subject to a trade-off between flow generation and translational motion of the swimmer. With hydrodynamic interactions, aggregation of swimmers becomes detrimental to mixing and pushers outperform pullers. Finally, we identify three distinct mixing stages and quantitatively investigate the dependence of mixing efficiency on system size.

The article is organized as follows. In Sec.~\ref{sec2}, we introduce our flow model, the method of mutual information, and the solution of advection-diffusion equation. In Sec.~\ref{sec3}, we present our results for non-motile stresslets with different spatial arrangements, optimal squirmer types with and without hydrodynamic interactions, and system-size effects. Section~\ref{sec4} concludes the paper.

%-------------------------------------------------------------------------

\section{Model}\label{sec2}
\subsection{The squirmer model}
At the microscale, inertia is negligible and, the fluid flow generated is governed by the incompressible Stokes equations:
\begin{equation}
    \label{stokeseqn}
-\mu\nabla^{2}\mathbf{\fvel}+\mathbf{\nabla}p=0,
\end{equation}
\begin{equation}
    \label{incompressible}
\nabla\cdot\mathbf{\fvel}=0,
\end{equation}
where $\mathbf{\fvel}$ is the velocity field, $\mu$ is the viscosity, and $p$ is the pressure.

The squirmer model was originally developed to describe ciliated organisms \cite{19}, and has since been extended to a wide range of microswimmers \cite{20, 21} and became a standard model of self-propelled, finite-size particles in a Stokes flow.
Here, we introduce 2D squirmers of radius $a$ in the case of non-deformable force-free circular particles with a tangential slip velocity $\mathbf{\fvel}^{s}$ in a polar coordinate system $(r,\phi)$ \cite{15}:
\begin{equation}
\label{surfacevel2}
 \fvel_{\phi}^{s}(a,\phi)=\sum_{n=1}^{\infty}B_n \sin{n\phi},
\end{equation}
where $B_n$ are constant coefficients that indicate the magnitude of the $n$th squirming mode and we have assumed the vanishing radial velocity, i.e., $v^s_r(a,\phi)=0$.
In the following, we use $\mathbf{x}_s$ to denote the location of the squirmer whose orientation is represented by a unit vector $\mathbf{e}$ (which defines the direction $\phi=0$). 

In the laboratory frame, the flow field around the squirmer in an unconfined fluid can be expressed by \cite{15}: 
\begin{equation}
\label{lab1}
\fvel_r(r,\phi)=\frac{B_1}{2}\frac{a^2}{r^2}\cos{\phi}+\sum_{n=2}^{\infty}\frac{n}{2}B_n \bigg[\frac{a^{n+1}}{r^{n+1}}-\frac{a^{n-1}}{r^{n-1}}\bigg]\cos{n\phi},
\end{equation}
\begin{equation}
\label{lab2}
\fvel_{\phi}(r,\phi)=\frac{B_1}{2}\frac{a^2}{r^2}\sin{\phi}+\sum_{n=2}^{\infty}\frac{1}{2}B_n \bigg[n\frac{a^{n+1}}{r^{n+1}}-(n-2)\frac{a^{n-1}}{r^{n-1}}\bigg]\sin{n\phi},
\end{equation}
where $\fvel_r$ and $\fvel_{\phi}$ are the radial and angular components of the flow velocity, respectively. The swimmer itself is translating with the velocity \begin{equation}
\label{translational}
\mathbf{V}=\frac{B_1}{2}\mathbf{e},
\end{equation}
which is determined by the $B_1$ mode. 

In the following, we retain only the first two modes, which dominate the far-field behavior and correspond to a source dipole and a force dipole, respectively. Accordingly, we set $B_n=0$ for all $n>2$.
We can re-write the flow field $\mathbf{\fvel}$ as the sum of the velocities from the first two modes: 
\begin{equation}
\label{twomodes}
\mathbf{\fvel}=\fvel_r \mathbf{e}_r+\fvel_{\phi} \mathbf{e}_{\phi} = \mathbf{\fvel}_{B_1} + \mathbf{\fvel}_{B_2},
\end{equation}
Here, $\mathbf{\fvel}_{B_1}$ is equivalent to a source dipole:
\begin{equation}
\label{B1}
\mathbf{\fvel}_{B_1}=-\frac{B_1 a^2}{2} \bigg(\frac{1}{2}\nabla_{0}^2\mathbf{G}(\vx-\mathbf{x}_s;\ve)\bigg),
\end{equation}
where $\mathbf{G}(\vx-\mathbf{x}_s;\ve)$ is the Stokeslet \cite{12} given in Eq.~\eqref{eq:stokeslet_green} of Appendix~\ref{app:singularities}, and the gradient $\nabla_0$ acts on the singularity position $\mathbf{x}_s$. This expression is equivalent to Eq.~\eqref{eq:source_dipole} of Appendix~\ref{app:singularities} with a prefactor defined by the squirmer model.
Also, $\mathbf{\fvel}_{B_2}$ corresponds to a stresslet contribution in the far field, with higher-order terms neglected:
\begin{equation}
\label{B2}
\mathbf{\fvel}_{B_2}=-B_2 a \bigg(\mathbf{e}\cdot\nabla_{0}\mathbf{G}(\vx-\mathbf{x}_s;\ve)\bigg) + \mathcal{O}(r^{-3}).
\end{equation}
Here, the first term is equivalent to Eq.~\eqref{eq:stresslet_axisymmetric} of Appendix~\ref{app:singularities}.
% Next, we consider the boundary condition of the Stokes equations Eq.~\eqref{stokeseqn}~\eqref{incompressible} on the surface of the swimmer:
% \begin{equation}
% \label{bc}
% \mathbf{\fvel}|_{r=a}=\mathbf{V}+\mathbf{\Omega}\times\mathbf{r}+\mathbf{\fvel}^s
% \end{equation}
% Here, the swimmer moves with the translational velocity $\mathbf{V}$ and the angular velocity $\mathbf{\Omega}$. In the absence of hydrodynamic interactions, we assume that the swimmer does not undergo rotation, so its angular velocity $\mathbf{\Omega}=0$. From the expressions above, the translational velocity of a squirmer in free space is therefore given by:
% \begin{equation}
% \label{translational}
% \mathbf{V}=\frac{B_1}{2}\mathbf{e}
% \end{equation}
% which is determined only by $B_1$ mode. 

The relative magnitude of $B_1$ and $B_2$ defines the squirmer parameter
\begin{equation}
\label{beta}
\beta=B_2/B_1, 
\end{equation}
which categorizes a squirmer as a stresslet, a pusher, a puller, and a source dipole. 
The squirmer with $|\beta|=\infty$ corresponds to a stresslet, which is equivalent to an axisymmetric force dipole. This limit is reached when $B_1=0$ and only the $B_2$ mode is present. In this case, the swimmer is non-motile ($\mathbf{V}=0$) in the absence of hydrodynamic interactions, and the resulting flow field $\mathbf{\fvel}$ is steady. 
Microorganisms such as \textit{Chlamydomonas reinhardtii} create thrust in the front, drawing fluid from the front and back of the body and ejecting it to the sides.
Microswimmers with this propulsion mechanism are classified as pullers with $\beta>0$ \cite{16}. 
Conversely, swimmers that generate thrust in the aft, such as \textit{E. coli}, are categorized as pushers with $\beta<0$ and draw fluid from the sides and expel it along the swimming direction. Neutral swimmers with $\beta=0$ are modeled as source dipoles, as exemplified by \textit{Paramecium}.  
The flow fields generated by stresslets, pushers, and pullers decay as $\mathbf{\fvel}\propto 1/r$ ($1/r^2$ in 3D), which is slower than the decay of source-dipole flow fields, $\mathbf{\fvel}\propto 1/r^2$ ($1/r^3$ in 3D). 

Without any external force or torque, the power expended by the swimmer and dissipated in the fluid can be expressed as \cite{9}:
\begin{equation}
\label{powerswim}
\mathcal{P}_A=-\int_{\mathcal{S}}\mathbf{f}_A\cdot\mathbf{v}^{s}d\mathbf{x}, 
\end{equation}
where $\mathcal{S}$ represents the boundary of the swimmer, and $\mathbf{f}_A=\mathbf{n}\cdot \bm{\sigma}_A$ is the traction force. Here, $\bm{\sigma}_A=-p\mathbf{I}+\mu\big(\nabla\mathbf{\fvel}^s+{\nabla\mathbf{\fvel}^s}^\top\big)$ with $\mathbf{I}$ being the identity matrix. 
%$\mathbf{I}$ denotes the stress field. 
By inserting the velocity fields given by Eqs.~(\ref{B1}) and (\ref{B2}) we get the dissipation rate
\begin{equation}
\label{PA}
\mathcal{P}_A=\mu \pi \big({B_1}^2+{B_2}^2\big).
\end{equation}

%----------------------------------------------------------------------

\subsection{Flow fields in a confined space}\label{flowfieldconfined}
We now consider the flow generated by a squirmer confined in a 2D square domain of side length $L$. To simplify the analysis while retaining confinement effects, we impose perfect-slip (stress-free) boundary conditions on all the walls. These read
\begin{align}
\label{boundary}
\fvel_x|_{x=0,L} &= 0, 
& \partial_x \fvel_y|_{x=0,L} &= 0, \\
\fvel_y|_{y=0,L} &= 0, 
& \partial_y \fvel_x|_{y=0,L} &= 0.
\end{align}
These conditions enforce zero normal velocity while allowing tangential slip along the boundaries, and enable a systematic construction of the Green’s function using the method of images.

We begin from the 2D Stokes equation with a point force $\mathbf{F}$ applied on $\mathbf{x}_s$,
\begin{align}
\label{stokespointforce}
-\mu\nabla^{2}\mathbf{\fvel}+\mathbf{\nabla}p=\mathbf{F}\delta(\mathbf{x}-\mathbf{x}_s), 
\qquad
\nabla\cdot\mathbf{\fvel}=0.
\end{align}
In free space, the solution is given by the 2D Stokeslet $\mathbf{G}(\mathbf{x}-\mathbf{x}_s;\mathbf{e})$ as described in Appendix~\ref{app:singularities}. The unit vector $\mathbf{e}$ represents the direction of the force $\mathbf{F}$. In confinement, we construct the solution by reflecting the singularity across the walls to enforce the boundary conditions. Repeated reflections generate an infinite periodic lattice of image singularities.

In practice, we compute the confined Green’s function in Fourier space. Expanding the velocity field in discrete Fourier modes $\mathbf{k}=\frac{\pi}{L}(k,s)$, one obtains the velocity in Fourier space (see Appendix~\ref{app:greens} for the derivation)
\begin{equation}
\label{fourierstokes}
\tilde{\mathbf{\fvel}}(\mathbf{k})=\frac{1}{\mu |\mathbf{k}|^2}\tilde{\mathbf{f}}(\mathbf{k})\cdot\bigg(\mathbf{I}-\frac{\mathbf{k}\mathbf{k}}{|\mathbf{k}|^2}\bigg).
\end{equation}
The confined Green’s function $\mathbf{G}^{(c)}$ is then obtained via inverse discrete Fourier transform, which is shown in Eq.~\eqref{eq:greensquare} of Appendix~\ref{app:greens}

Higher-order singularities corresponding to the squirmer modes are obtained by differentiation of the confined Green’s function with respect to the singularity position $\mathbf{x}_s$. The leading contributions retained here are the stresslet 
\begin{equation}
\label{stresslet}
\mathbf{\fvel}_{B_2}^{(c)}(\mathbf{x})=-B_2 a \bigg(\mathbf{e}\cdot\nabla_{0}\mathbf{G}^{(c)}\bigg),
\end{equation}
and the source dipole
\begin{equation}
\label{sourcedipole}
\mathbf{\fvel}_{B_1}^{(c)}(\mathbf{x})=-\frac{B_1 a^2}{2} \bigg(\frac{1}{2}\nabla_{0}^2\mathbf{G}^{(c)}\bigg).
\end{equation}
To ensure numerical stability and represent a finite-sized swimmer, we therefore introduce a smooth spectral regularization
\begin{equation}
\tilde{\mathbf{\fvel}}^{(r)}(\mathbf{k})
=
\tilde{\mathbf{\fvel}}(\mathbf{k})
\ee^{-\varepsilon |\mathbf{k}|/\pi},
\end{equation}
where $\varepsilon = 2a$ represents the effective swimmer size. 
This exponential damping suppresses high-wavenumber modes, removes numerical oscillations, and effectively regularizes the singular flow near the swimmer position.
The regularised velocity field of a squirmer confined in the square is then given by
\begin{equation}
\mathbf{\fvel}(\mathbf{x})=\mathbf{\fvel}^{(r)}_{B_1}(\mathbf{x})+\mathbf{\fvel}^{(r)}_{B_2}(\mathbf{x}). 
\end{equation}
The full derivation is shown in Appendix~\ref{app:squirmer}. Figure~\ref{fig:flowfield} shows the streamlines of four types of swimmers placed at the center of the square domain: the source dipole, the pusher, the puller, and the stresslet.
\begin{figure*}	
\centering
\includegraphics[width =\textwidth]{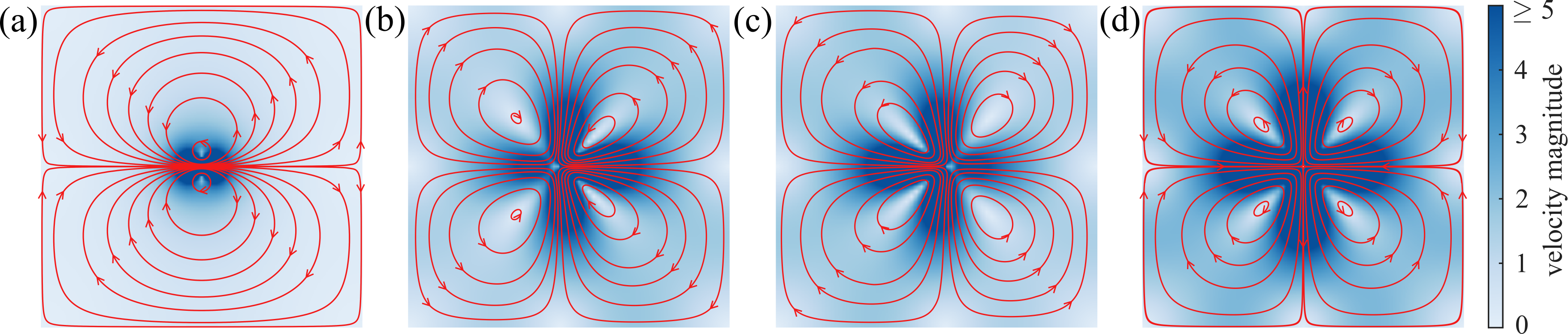} 
\caption{Representative streamlines of the regularized confined velocity field generated by a single squirmer placed at the center of a 2D square domain with perfect-slip boundaries. The four swimmer types are compared under the same total dissipation constraint: (a) source dipole, (b) pusher, (c) puller, and (d) stresslet, each oriented along $\theta=0$. Color indicates the velocity magnitude.}
\label{fig:flowfield}		
\end{figure*}

The flow generated by multiple swimmers is obtained by superposition. Unless stated otherwise, swimmers move with translational velocity as in Eq.~\eqref{translational}, and follow straight trajectories between wall reflections as stated in Appendix~\ref{app:squirmer}. Hydrodynamic interactions between swimmers are incorporated where explicitly indicated later in Sec.~\ref{sec3}.

The construction in this section provides a numerically stable representation of the confined squirmer flow field, which will be used to solve the advection–diffusion (Fokker–Planck) equation for tracer dynamics.

\subsection{Mutual information and Fokker-Planck equation}
Having established the flow model, we now introduce the measure used to quantify mixing efficiency. Fluid mixing can be characterized by loss of information across the entire system. Following our previous work \cite{8}, we introduce mutual information, a concept from information theory \cite{infobook}, as a universal assumption-free measure of mixing efficiency. Here, we define the mutual information between the initial position $\mathbf{x}_{0}$ and the final position $\mathbf{x}_{t}$ of a fluid particle as
\begin{equation}\label{mutualmain}
    I[\mathbf{x}_{0};\mathbf{x}_{t}]=S[\mathbf{x}_{t}]-S[\mathbf{x}_{t}|\mathbf{x}_{0}],
\end{equation}
where 
\begin{equation}
\label{entropy1}
S[\mathbf{x}_{t}] = -\int P(\mathbf{x},t)\log{P(\mathbf{x},t)} \,d\mathbf{x},
\end{equation}
is the Shannon entropy of the position variable at time $t$, and 
\begin{equation}
\label{entropy2}
S[\mathbf{x}_{t}|\mathbf{x}_{0}] = -\int{P(\mathbf{x}_{0})\left[\int{P(\mathbf{x},t|\mathbf{x}_{0})\log P(\mathbf{x},t|\mathbf{x}_{0})}d\mathbf{x}\right]}d\mathbf{x}_{0},
\end{equation}
is the conditional entropy with the knowledge of the initial position. Here, $P(\mathbf{x},t)=\int P(\mathbf{x},t\vert \mathbf{x}_0) P(\mathbf{x}_0)d\mathbf{x}_0$ is the distribution of position variable $\mathbf{x}$ at time $t$, $P(\mathbf{x},t|\mathbf{x}_{0})$ is the conditional probability with the knowledge of the initial position of the particle, and $P(\mathbf{x}_0)$ is the distribution of initial positions.  This initial distribution $P(\mathbf{x}_0)$ determines the statistical weight that is given to the mixing efficiency in different regions of the fluid. It should not be confused with any imposed pattern in the fluid to be ``erased'' by mixing, as frequently used in other mixing efficiency criteria --- mutual information does not require any assumptions about the initial spatial distribution of the fluid components to be mixed. Throughout this work, we adopt a homogeneous weight $P(\mathbf{x}_{0})= 1/V$, where $V=L^2$ is the total area of the domain. For incompressible flows, the uniform distribution is stationary, implying $P(\mathbf{x},t)= 1/V$ at all times. In this form, mixing is entirely governed by the spreading of the conditional probability density $P(\mathbf{x},t|\mathbf{x}_{0})$.
  
Mutual information, as defined above, measures the information that the final position (at time $t$) of the fluid particle contains about its initial position. During the mixing process, the conditional distribution broadens and the dependence on the initial position weakens, so $I[\mathbf{x}_{0};\mathbf{x}_{t}] \equiv I(t)$ monotonically decreases with time. Perfect mixing corresponds to $I(t)=0$, where the final and initial positions are statistically independent.

The transport of tracers is described by the advection–diffusion (Fokker–Planck) equation
\begin{equation}
     \label{FP}
     \partial_{t} P + \mathbf{\fvel}\cdot\nabla P=D\nabla^2 P,
\end{equation}
for a probability density $P(\mathbf{x},t)$ evolving under the confined flow field $\mathbf{\fvel}(\mathbf{x},t)$ and the diffusion constant $D$. The no-flux condition implies $\mathbf{n}\cdot\nabla P(\mathbf{x},t)=0$ at all boundaries. The conditional probability density $P(\mathbf{x},t|\mathbf{x}_{0})$ in 
Eq.~\eqref{entropy2} is obtained as the Green’s function of the advection–diffusion equation with the initial condition $\mathbf{x}=\mathbf{x}_0$ at $t=0$.

\begin{figure*}	
\centering
\includegraphics[width =\textwidth]{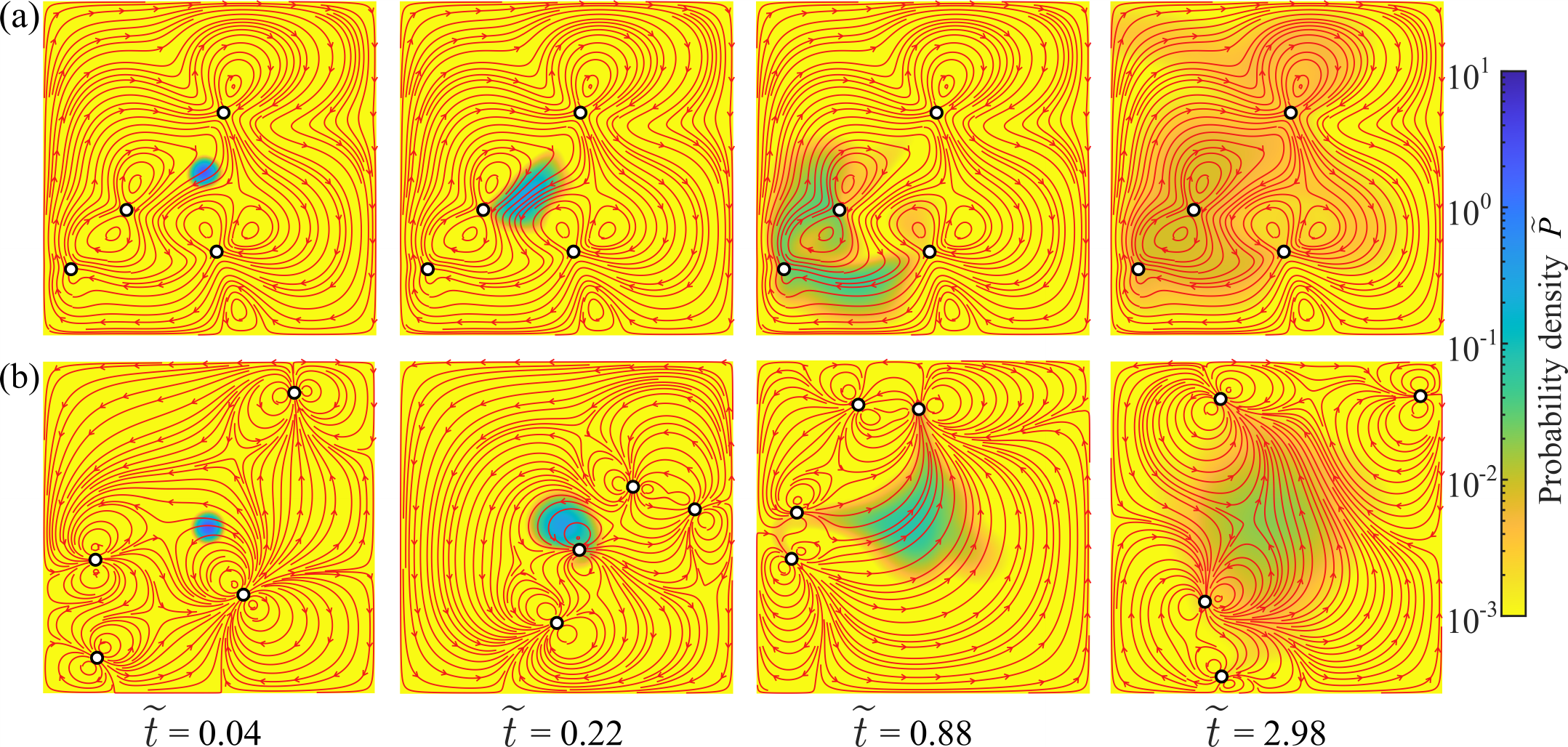} 
\caption{Time evolution of the probability density $\ndP(\mathbf{x},\ndt|\mathbf{x}_{0})$ with an initial position $\mathbf{x}_{0}$ in the middle of the square domain. The flow fields are shown with red arrows. The empty circles represent the singularity positions of swimmers. (a) System with $4$ non-motile stresslets ($\widetilde{B}_2=85$). (b) System with $4$ source dipoles ($\widetilde{B}_1=85$). In both systems, the swimmers are randomly located and oriented at $\ndt=0$. The hydrodynamic interactions are not considered. Animated versions of both panels are provided as Movie~1 (a) and Movie~2 (b) in the Supplemental Material~\cite{SM}.}
\label{fig:movie}		
\end{figure*}

In the following, we nondimensionalize all variables using the length scale $L_s=L/20 \sim a$ and the time scale $T_s=L_s^2/D$ such that $\ndt=t/T_s$, $\ndL=L/L_s$, etc. In order to numerically determine the conditional entropy $S[\mathbf{x}_{t}|\mathbf{x}_{0}]$ in Eq.~\eqref{mutualmain}, we use a pseudo-spectral method to calculate $P(\mathbf{x},t|\mathbf{x}_{0})$ for any initial position $\mathbf{x}_{0}$. Two examples of the nondimensionalized solution are shown in Fig.~\ref{fig:movie}: (a) with the stationary flow field of $4$ non-interacting stresslets that are non-motile; (b) with the evolving flow field of $4$ non-interacting source dipoles that move at constant speed. The conditional entropy $S[\mathbf{x}_{t}|\mathbf{x}_{0}]$ is obtained by integrating $P(\mathbf{x},t|\mathbf{x}_{0})\log P(\mathbf{x},t|\mathbf{x}_{0})$ over $\widetilde{x}_0$, $\widetilde{y}_0$, $\widetilde{x}$ and $\widetilde{y}$, which is needed to determine the mutual information $I(t)$.

\section{Results}\label{sec3}
We now investigate the mixing induced by microswimmers using the model introduced in Sec.~\ref{sec2}. Swimmers are modeled using the squirmer representation, which allows us to distinguish several types of limiting flow, including stresslets, pushers, pullers, and source dipoles. The mixing efficiency is quantified by the decay of mutual information obtained from the solution of the advection–diffusion equation.

Our results address three key aspects of swimmer-induced mixing. First, we examine how the spatial arrangement of swimmers affects mixing by studying non-interacting stresslets in different configurations. Second, we compare the mixing efficiency generated by different swimmer types under the constraint of fixed total dissipation, and investigate how hydrodynamic interactions modify the optimal swimmer type. Finally, we analyze the time evolution of mixing in systems of different sizes, which reveals distinct regimes governed by diffusion, advection, and relaxation.

\subsection{Spatial arrangement of non-interacting stresslets}
To isolate the influence of swimmer organization on mixing, we first consider systems of identical stresslets with strength $\widetilde{B}_2$ confined in a square domain of side length $\widetilde{L}$. Hydrodynamic interactions between swimmers are neglected, so that the steady velocity field is obtained as a linear superposition of the flows generated by individual stresslets. In this setting, the spatial arrangement and orientation of swimmers determine the structure of the velocity field and therefore the efficiency of mixing.

\begin{figure*}[t]
\centering
   \includegraphics[width = \textwidth]{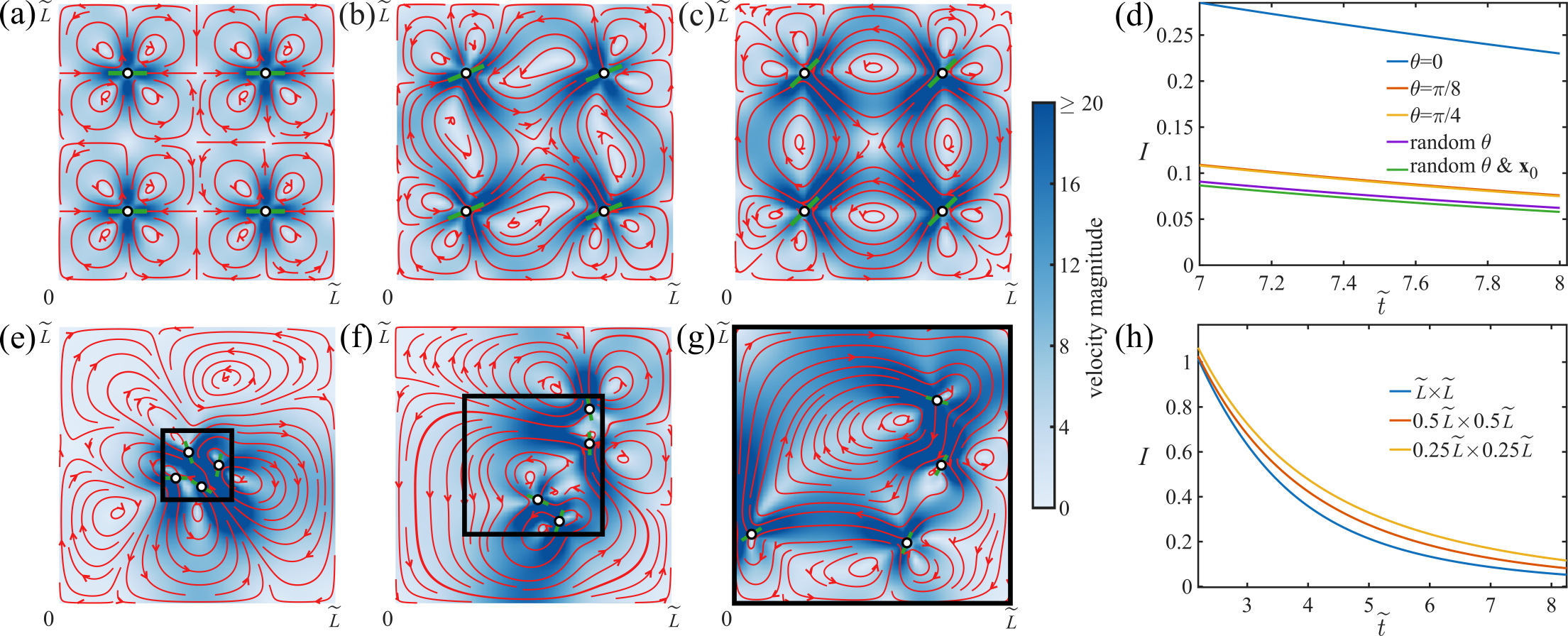}
\caption{Velocity fields and mixing efficiency for systems of four non-interacting stresslets in a square domain of side length $\widetilde{L}$. (a–c) Velocity fields for swimmers located at ordered positions with identical orientation angles $\theta=0,\pi/8,$ and $\pi/4$, respectively. Green rods indicate the force-dipole representation of each stresslet. (d) Corresponding evolution of mutual information $I(\ndt)$ for the configurations in (a–c), compared with systems with ordered positions but random orientations ($\theta$) and systems with both random orientations and positions ($\theta, \mathbf{x}_0$). (e–g) Velocity fields for swimmers randomly oriented and randomly located within a central square region of side length $l'=0.25\widetilde{L}, 0.5\widetilde{L},$ and $\widetilde{L}$, respectively. (h) Corresponding mutual information evolution for the systems in (e–g). All swimmers have stresslet strength $\widetilde{B}_2=85$ and radius $\widetilde{a}=\frac{3}{64}\widetilde{L}$. 
}
\label{fig:spatial}		
\end{figure*}

Figure~\ref{fig:spatial}(a-d) shows examples of velocity fields generated by four stresslets arranged in ordered positions within the square domain. The swimmers share the same orientation characterized by the unit vector $\mathbf{e}=(\cos{\theta},\sin{\theta})$. Panels (a–c) illustrate three configurations with orientation angles $\theta=0,\pi/8,$ and $\pi/4$, respectively. When $\theta=0$, the symmetry of the configuration divides the flow field into four domains without advective flow between them. As the orientation angle increases, this symmetry is broken. The vortices extend across larger portions of the system and connect neighboring regions of the flow, which enhances advective transport, as shown in Fig.~\ref{fig:spatial}(b,c).
The corresponding evolution of mutual information is shown in panel (d), where the configurations in panels (a–c) are compared with systems having random swimmer orientations and with systems having both random positions and orientations across the entire domain. For the random configurations, the mutual information is averaged over several realizations of the swimmer distribution.

Consistent with physical intuition, disorder in swimmer orientations accelerates mixing. Systems with ordered positions but random orientations exhibit a faster decay of mutual information than the symmetric configurations shown in panels (a–c). Introducing positional disorder in addition to orientational disorder still produces slightly faster decay, although the difference between these two disordered cases is relatively small. In these systems, the superposition of stresslet flows generates more irregular velocity fields that enhance the stretching and dispersion of tracer distributions. 

We next examine the influence of swimmer aggregation. Figure~\ref{fig:spatial}(e-h) shows systems in which swimmers are randomly oriented and randomly positioned within a square region of side length $l'$ located at the center of the domain. Panels (e-g) correspond to $l'=0.25\widetilde{L}, 0.5\widetilde{L},$ and $\widetilde{L}$, respectively. When swimmers are strongly clustered ($l'=0.25\widetilde{L}$), the induced velocity field is concentrated near the center of the domain and has little influence on the outer regions, resulting in relatively slow mixing. As the spatial extent of the swimmer distribution increases, the flow field becomes more uniformly distributed across the system and the transport of tracers is enhanced. This trend is reflected in the decay of mutual information shown in panel (h), where larger values of $l'$ correspond to faster mixing.

Overall, these results demonstrate that the spatial organization of microswimmers plays a crucial role in determining mixing efficiency. Highly symmetric configurations and strong aggregation of swimmers tend to suppress large-scale transport, whereas disorder in swimmer orientations and positions, as well as broader spatial dispersion of swimmers, generate more complex flow structures that enhance fluid mixing.

\begin{figure*}
\centering
   \includegraphics[width = \textwidth]{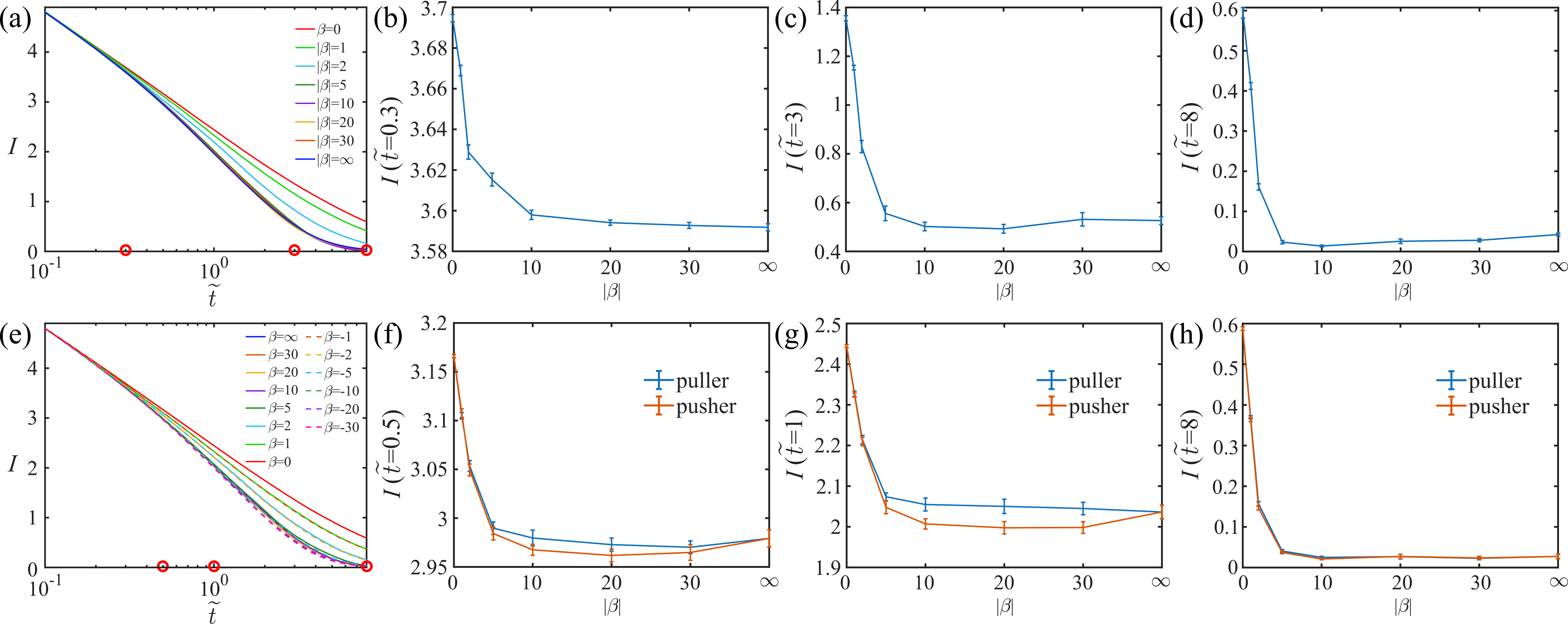}
\caption{Mixing efficiency for different swimmer types characterized by the squirmer parameter $\beta$ under the constraint of fixed total dissipation $\mathcal{P}$. (a) Time evolution of mutual information $I(\ndt)$ for systems with different values of $|\beta|$ in the absence of hydrodynamic interactions. The red circles indicate the time moments when the dependence is shown in (b-d) accordingly. (b–d) Mutual information as a function of $|\beta|$ at fixed times, $\ndt=0.3, 3,$ and $8$, respectively, showing that the mixing efficiency depends non-monotonically on swimmer type and that the minimum mutual information occurs at finite $|\beta|$. (e) as in (a) but with hydrodynamic interactions between swimmers. (f–h) Mutual information as a function of $|\beta|$ at fixed times, $\ndt=0.5, 1,$ and $8$, respectively, for the interacting system. In this case, pushers and pullers exhibit different mixing efficiencies because hydrodynamic interactions modify the swimmer trajectories. All systems contain $N=4$ identical swimmers and satisfy the same dissipation constraint $\widetilde{\mathcal{P}}_A=10^4 \widetilde{\mu} \pi$. All results are given by the average of 20 samples with random locations and orientations of swimmers. The error bars represent SEM (standard error of the mean).   
}
\label{fig:dissipation}		
\end{figure*}

\subsection{Optimal swimmer type with fixed dissipation}\label{optimalspecies}

We next compare the mixing efficiency generated by different types of microswimmers under the constraint of a fixed total dissipation rate. The swimmer type is characterized by the squirmer parameter $\beta=\widetilde{B}_2/\widetilde{B}_1$, which distinguishes stresslets ($|\beta|\rightarrow\infty$), pushers ($\beta<0$), pullers ($\beta>0$), and source dipoles ($\beta=0$). For a circular squirmer in a viscous fluid, the power dissipated by a single swimmer is determined by Eq.~\eqref{PA}. We carry out the comparison between swimmer types by fixing the total dissipation rate
\begin{equation}
\label{totaldissi}
\mathcal{P}=N {\widetilde{\mathcal{P}}_A} =N\widetilde{\mu} \pi \big({\widetilde{B}_1}^2+{\widetilde{B}_2}^2\big),
\end{equation}
where $N$ is the number of swimmers and $\widetilde{\mu}$ is the viscosity. Here, we keep $N=4$ and ${\widetilde{B}_1}^2+{\widetilde{B}_2}^2=10^4$ for all species. Varying the swimmer type through the parameter $\beta$ therefore corresponds to choosing different pairs $(\widetilde{B}_1, \widetilde{B}_2)$ that satisfy the same dissipation constraint.

We first consider the case where hydrodynamic interactions between swimmers are neglected. The swimmer trajectories are then prescribed and independent of the flows generated by other swimmers and walls. Figure~\ref{fig:dissipation}(a) shows the evolution of mutual information $I(\ndt)$ for systems with different values of $|\beta|$. Pushers and pullers with equal magnitude of $\beta$ produce identical mixing efficiency, reflecting the time-reversal symmetry of mutual information \cite{8}. Consequently, the decay of mutual information depends only on $|\beta|$.

The dependence of mixing efficiency on $|\beta|$ becomes clearer when mutual information is examined at fixed times, as shown in Fig.~\ref{fig:dissipation}(b–d). At early stage as in (b), mutual information decreases monotonically with increasing $|\beta|$, indicating that swimmers generating stronger dipolar flows enhance stirring of the fluid. This dependence can be understood from the spatial uniformity of the velocity field. The flow field generated by a stresslet decays most slowly, scaling as $\mathbf{\fvel}\propto 1/r$, whereas that of a source dipole decays more rapidly, following $\mathbf{\fvel}\propto 1/r^2$. Because the local reduction of mutual information at short times depends logarithmically on the dissipation density \cite{10}, a distribution of dissipation that is spatially more uniform leads to a higher overall mixing efficiency. 

At later stages of the mixing process as in (c,d), the dependence becomes non-monotonic. The minimum mutual information occurs at a finite value of $|\beta|$ rather than in the stresslet limit. This behavior arises because stresslets with $|\beta|\rightarrow\infty$ are non-motile in the absence of hydrodynamic interactions ($\widetilde{B}_1=0$), whereas swimmers with finite $|\beta|$ translate through the domain and assure a spatially more uniform effect. 
This indicates that although the stresslet component promotes mixing through longer-ranged flows, swimmer translation also contributes to mixing by spreading the areas with strong shear rates across the system.

We next include hydrodynamic interactions between swimmers. In this case, the motion of each swimmer is influenced by the flow generated by the others and is determined using Fax\'{e}n's laws for the translational and rotational velocities (see Appendix~\ref{app:hi} for more details). The wall effect on the motion of swimmers is neglected. The resulting mixing dynamics are shown in Fig.~\ref{fig:dissipation}(e–h). In contrast to the non-interacting case, pushers with the same magnitude of $\beta$ now exhibit faster decay of mutual information than pullers during the intermediate stage of the mixing process. This is in agreement with the study of long-range diffusivity in dilute solutions \cite{33}.  The effect can be explained with the tendency of pullers to form aggregates or show polar alignment \cite{Alarcon.Pagonabarraga2013,Oyama.Yamamoto2016,Zantop.Stark2022}, which reduces the mixing efficiency as shown above.
Moreover, the dependence of mixing efficiency on $|\beta|$ becomes smoother and the optimal swimmer type shifts toward larger values of $|\beta|$. These results suggest that swimmer-induced dipolar flows provide the dominant mechanism for stirring the fluid, while swimmer motion and hydrodynamic interactions determine the optimal balance between propulsion and flow generation for efficient mixing.

\begin{figure*}
\centering
   \includegraphics[width = \textwidth]{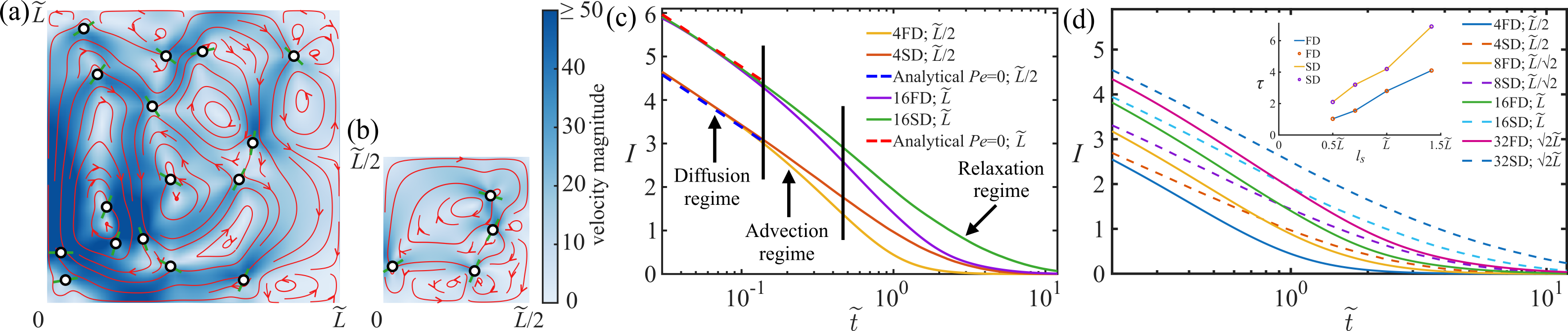}
\caption{Mixing in systems with different square sizes at the fixed number density of non-interacting swimmers. 
(a) Example of the flow velocity field (red arrows) generated by 16 randomly distributed stresslets (empty circles, with green segments indicating their orientation) in a square domain with side length $l_s=\widetilde{L}$.  
(b) Same as (a) but with 4 stresslets in a smaller square domain with $l_s=\widetilde{L}/2$, corresponding to the same swimmer number density. 
(c) Evolution of mutual information $I(\ndt)$ for stresslets (FD) and source dipoles (SD) in systems with $l_s=\widetilde{L},\widetilde{L}/2$. 
The dashed lines show the analytical approximation for pure diffusion ($Pe=0$),
$I(\ndt)=-\log(4\pi \ndt/A)-1$, where $A=l_s^2$ is the area of the square domain. 
(d) Evolution of mutual information $I(\ndt)$ at later times for stresslets and source dipoles in systems with $l_s=\widetilde{L}/2$, $\widetilde{L}/\sqrt{2}$, $\widetilde{L}$, and $\sqrt{2}\widetilde{L}$. 
The inset shows the dependence between relaxation time $\tau$, extracted from the exponential decay $I\propto\exp(-\ndt/\tau)$, and side length $l_s$. The number density of swimmers in all systems is fixed at $n=16/\widetilde{L}^2$. The stresslets have a magnitude $\widetilde{B}_2=100$ and source dipoles $\widetilde{B}_1=100$.  
}
\label{fig:3stage}
\end{figure*}

\subsection{Mixing regimes and system-size dependence}
In this section we study different stages of the mixing process in systems with different square side lengths $l_s$ at a fixed number density of non-interacting swimmers. Examples of the flow fields generated by randomly distributed stresslets in square domains with $l_s=\widetilde{L}, \widetilde{L}/2$ are shown in Fig.~\ref{fig:3stage}(a,b). Although the local flow structures remain similar, increasing the system size enlarges the spatial region over which tracers must be transported. The swimmer number density in all systems considered here is fixed at $n=16/\widetilde{L}^2$.

Figure~\ref{fig:3stage}(c) shows the evolution of mutual information $I(\ndt)$ for stresslets (FD) and source dipoles (SD) for two sizes of the fluid domain. 
All mixing processes undergo three stages: diffusion regime, advection-diffusion regime, and the final relaxation stage. Overall, mixing induced by source dipoles is less efficient than that induced by stresslets, consistent with the results in Sec.~\ref{optimalspecies}. For each swimmer type, particles in larger systems retain more information than those in smaller systems, and the mutual information therefore remains higher throughout the mixing process.
Quantitatively, we compare numerical results with the analytical approximation for pure diffusion ($Pe=0$) in free space \cite{8},
$I(\ndt)= -\log{(4\pi \ndt/A)}-1$, where $A=l_s^2$ denotes the area of the square domain and the P\'{e}clet number $Pe=Ua/D$ is based on the characteristic speed $U$ of the swimmer-generated flow rather than the swimmer's translational velocity. The agreement between the analytical prediction (dashed lines) and the numerical results (solid lines) indicates that the initial stage of mixing is diffusion-dominated.

To further examine the intermediate stage and the final relaxation, we analyze the evolution of mutual information at later times by comparing more cases with different $l_s$, as shown in Fig.~\ref{fig:3stage}(d). After the diffusion-dominated stage, the influence of advection becomes evident for $\ndt<1$. In this regime, for a fixed swimmer density, the decay rate of $I$ is nearly independent of the system size. The difference between mutual information in systems with areas $A_1$ and $A_2$ remains approximately $\Delta I\sim\log(A_1/A_2)$, consistent with the scaling expected from the diffusion approximation, indicating that advection accelerates the mixing process in a way that is largely independent of the system size.
This behavior can also be understood as follows. During the early (diffusion-dominated) and intermediate (advection-enhanced) stages, most tracer particles do not interact with the system boundaries. In contrast, during the final relaxation stage, tracer particles reach the boundaries and the system size becomes the dominant factor controlling the mixing dynamics.

To characterize the final relaxation stage, we analyze the relaxation time constant $\tau$ for the processes shown in Fig.~\ref{fig:3stage}(d), where $\tau$ is defined through $I\propto \exp(-\ndt/\tau)$. The values of $\tau$ extracted from the long-time behavior of the curves are summarized in the inset of Fig.~\ref{fig:3stage}(d). We find that the dependence of $\tau$ on $l_s$ is weaker than $\propto l_s^2$, which is expected for a process that can be described as an effective diffusion in confined space. The deviation can be explained with the fact that larger confinement not only increases the length over which the particle distribution needs to be equilibrated, but also widens the distance over which the advective flows act.

\section{Conclusion}\label{sec4}
In this work, we investigated fluid mixing induced by microswimmers using mutual information as a global measure of mixing efficiency. This information-theoretic approach characterizes mixing without imposing any specific initial concentration pattern, providing a universal and physically transparent description of mixing processes in active fluids. Based on the two-dimensional squirmer model, we constructed the confined flow field in a square domain using the method of images and quantified the mixing dynamics by solving the advection–diffusion (Fokker–Planck) equation. 

Our results reveal several factors governing swimmer-induced mixing. First, the spatial organization of swimmers: highly symmetric configurations and strong aggregation generate localized flow structures that limit large-scale transport, whereas orientational and positional disorder produce more irregular velocity fields that significantly enhance mixing. Second, under the constraint of fixed energy dissipation, the mixing efficiency depends non-monotonically on the squirmer parameter $|\beta|$. In the absence of hydrodynamic interactions, the optimal mixing efficiency occurs at a finite $|\beta|$, reflecting a balance between long-range dipolar flows and swimmer translation. When hydrodynamic interactions are included, pushers outperform pullers, and the optimal swimmer type shifts toward larger $|\beta|$, demonstrating the importance of collective hydrodynamic effects. Third, by comparing systems of different sizes at fixed swimmer density, we identified three characteristic regimes of the mixing dynamics: an initial diffusion-dominated regime, an intermediate regime where advection accelerates the decay of mutual information, and a final relaxation stage controlled primarily by the system size.

The present study demonstrates that mutual information provides a powerful measure for quantifying mixing in active and microscale flows. It offers a universal and assumption-free method for quantifying mixing in complex dynamical systems. The approach introduced here is broadly applicable to a variety of biological and active-matter systems in which fluid transport is driven by the motion of microscopic agents, such as ciliary flows, microbial suspensions, and intracellular transport processes. In this context, our results provide quantitative insight into how swimmer type, spatial organization, and collective hydrodynamic interactions shape mixing efficiency, and thereby contribute to a deeper understanding of transport phenomena in active fluids and other complex flow environments. Our framework for quantification of the energy dissipation (or entropy production) involved in the mixing process can be further extended by using appropriate measures from stochastic thermodynamics, such as the recently developed many-body generalization of the Harada-Sasa relation \cite{Golestanian2025}.

\begin{acknowledgments}
We acknowledge support from the Max Planck Center Twente for Complex Fluid Dynamics, the Max Planck Society, the Max Planck School Matter to Life, and the MaxSynBio Consortium, which are jointly funded by the Federal Ministry of Education and Research (BMBF) of Germany.
Y.H.\ acknowledges support from the Japan Science and Technology Agency (JST) CREST (Grant No.\ JPMJCR25Q1). A.V. acknowledges support from the Slovenian Research and Innovation Agency (Grants No.\ P1-0099 and J1-60009).
\end{acknowledgments}

\clearpage

\appendix
\section{Far-field singularities of a microswimmer in free space}
\label{app:singularities}

\subsection{Fundamental singularities}

We first summarize the far-field singularities used to construct the flow generated by a microswimmer. In two-dimensional incompressible Stokes flow, the velocity field generated by a point force \(\vF=(F_x,F_y)\) applied at \(\mathbf{x}_s=(x_s,y_s)\) satisfies
\begin{equation}
-\mu\nabla^{2}\vv+\nabla p=\vF\delta(\vx-\mathbf{x}_s),
\label{eq:stokeslet_eqn}
\end{equation}
with \(\nabla\cdot\vv=0\). Let \(\vF=\alpha\ve\), where \(\ve=(e_1,e_2)\) is a unit vector. The free-space Stokeslet solution is
\begin{equation}
\vv(\vvr)=\frac{\alpha}{4\pi\mu}\mathbf{G}(\vvr;\ve),
\label{eq:stokeslet}
\end{equation}
where
\begin{equation}
\mathbf{G}(\vvr;\ve)
=-\ve\log\frac{r}{c}+\frac{\ve\cdot\vvr}{r^2}\vvr,
\label{eq:stokeslet_green}
\end{equation}
\(\vvr=\vx-\mathbf{x}_s\), and \(r=|\vvr|\). The constant \(c\) reflects the fact that the two-dimensional Stokeslet is defined only up to an additive constant velocity, which is related to the Stokes paradox.

Although a force-free swimmer cannot be represented by a single Stokeslet, derivatives of the Stokeslet generate the leading singularities appearing in the far-field expansion of microswimmer flows. The Stokeslet dipole is
\begin{equation}
\begin{aligned}
&\mathbf{G}_d(\vvr;\vd,\ve)\\
&=\vd\cdot\gradzero\mathbf{G}(\vvr;\ve)\\
&=\left(\frac{2(\ve\cdot\vvr)(\vd\cdot\vvr)}{r^4}-\frac{\ve\cdot\vd}{r^2}\right)\vvr
+\frac{-(\ve\cdot\vvr)\vd+(\vd\cdot\vvr)\ve}{r^2}.
\end{aligned}
\label{eq:stokeslet_dipole}
\end{equation}
The symmetric part is
\begin{align}
\mathbf{G}_d^{\mathrm{sym}}(\vvr;\vd,\ve)
&=\frac{1}{2}\left[\mathbf{G}_d(\vvr;\vd,\ve)+\mathbf{G}_d(\vvr;\ve,\vd)\right]\nonumber\\
&=\left(\frac{2(\ve\cdot\vvr)(\vd\cdot\vvr)}{r^4}-\frac{\ve\cdot\vd}{r^2}\right)\vvr.
\label{eq:stresslet_general}
\end{align}
For \(\vd=\ve\), this gives the axisymmetric stresslet

\begin{equation}
\mathbf{G}_d^{\mathrm{sym}}(\vvr;\ve,\ve)
=\ve\cdot\gradzero\mathbf{G}(\vvr;\ve)
=\left(\frac{2(\ve\cdot\vvr)^2}{r^4}-\frac{1}{r^2}\right)\vvr.
\label{eq:stresslet_axisymmetric}
\end{equation}
The antisymmetric part is
\begin{align}
\mathbf{G}_d^{\mathrm{antisym}}(\vvr;\vd,\ve)
&=\frac{1}{2}\left[\mathbf{G}_d(\vvr;\vd,\ve)-\mathbf{G}_d(\vvr;\ve,\vd)\right]\nonumber\\
&=\frac{-(\ve\cdot\vvr)\vd+(\vd\cdot\vvr)\ve}{r^2}.
\label{eq:rotlet_general}
\end{align}
Taking \(\vd=\ve^{\perp}\), where \(\ve^{\perp}\cdot\ve=0\), yields the rotlet
\begin{equation}
\mathbf{G}_d^{\mathrm{antisym}}(\vvr;\ve^{\perp},\ve)
=\frac{-(\ve\cdot\vvr)\ve^{\perp}+(\ve^{\perp}\cdot\vvr)\ve}{r^2}.
\label{eq:rotlet}
\end{equation}

\begin{figure}[ht]
\centering
\includegraphics[width=0.75\columnwidth]{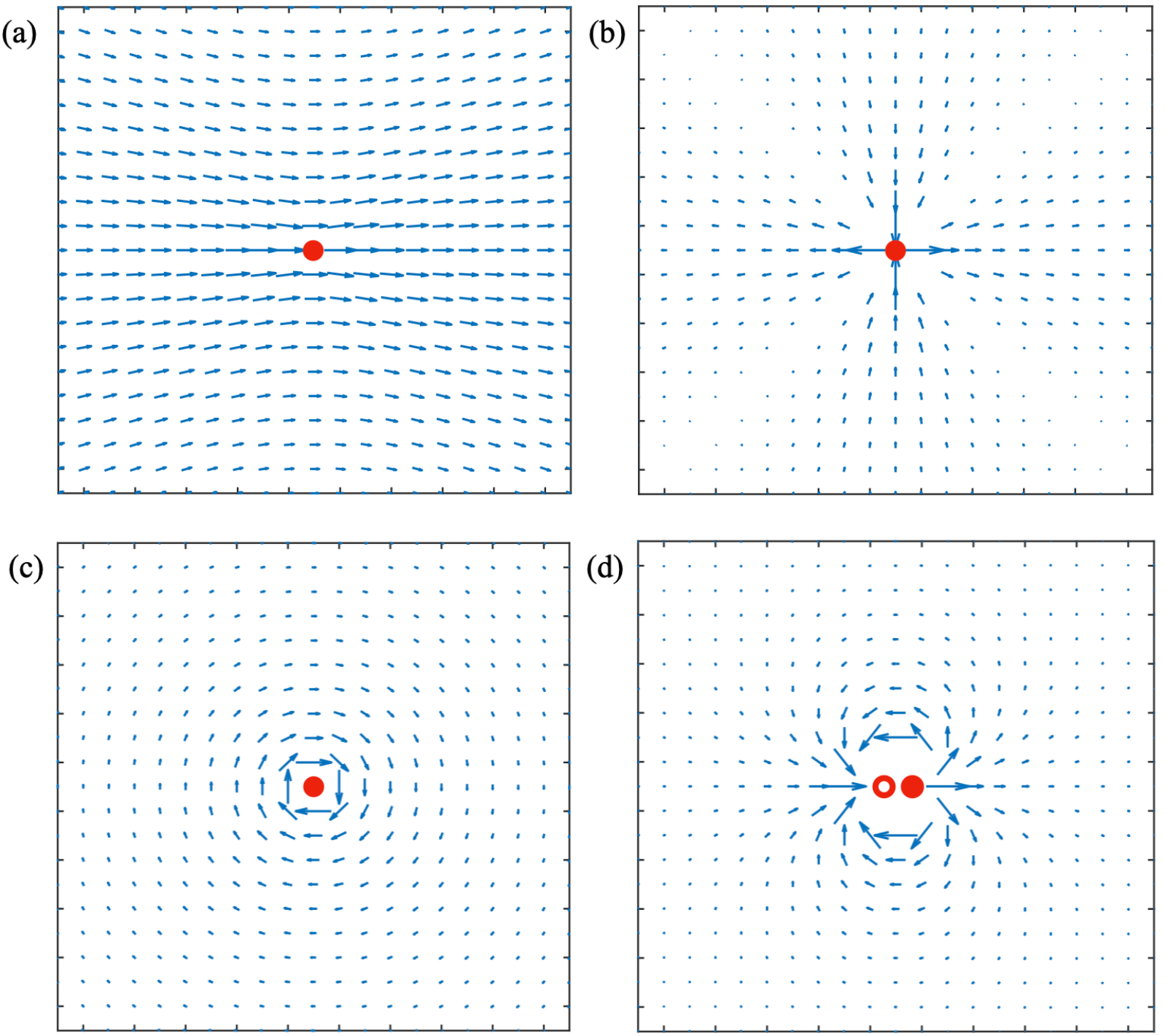}
\caption{Velocity fields of fundamental singularities in two-dimensional unconfined fluids: (a) Stokeslet, (b) stresslet, (c) rotlet, (d) source dipole.}
\label{fig:free_space_singularities}
\end{figure}

The source dipole is obtained from the Laplacian of the Stokeslet,
\begin{equation}
\mathbf{D}(\vvr;\ve)
=\frac{1}{2}\gradzero^2\mathbf{G}(\vvr;\ve)
=\ve\cdot\left(\frac{\vI}{r^2}-\frac{2\vvr\vvr}{r^4}\right),
\label{eq:source_dipole}
\end{equation}
where \(\vI\) is the identity matrix.
See Fig.~\ref{fig:free_space_singularities} for the velocity fields induced by the fundamental singularities discussed above.

\section{Confined Green's function in a square domain}
\label{app:greens}

\subsection{Perfect-slip boundary conditions and image construction}

We next construct the flow field of a singularity confined in a square domain of side length \(L\). The square is bounded by perfect-slip walls, for which
\begin{equation}
 v_x|_{x=0,L}=0,
\qquad
 \frac{\partial v_y}{\partial x}\bigg|_{x=0,L}=0,
\label{eq:perfect_slip_bc_x}
\end{equation}
\begin{equation}
 v_y|_{y=0,L}=0,
\qquad
 \frac{\partial v_x}{\partial y}\bigg|_{y=0,L}=0.
\label{eq:perfect_slip_bc_y}
\end{equation}
These conditions enforce zero normal velocity while allowing tangential slip along the boundaries.

\begin{figure}[b]
\centering
\includegraphics[width=0.45\columnwidth]{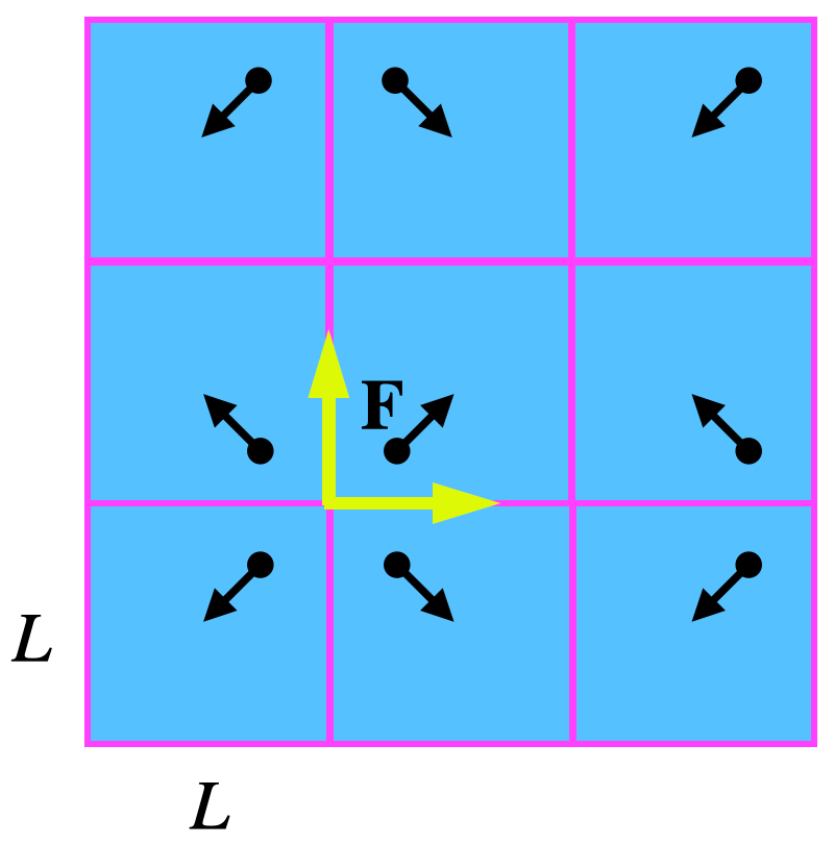}
\caption{Image construction for solving the Stokeslet in a square with perfect-slip boundary conditions.}
\label{fig:image_method}
\end{figure}

To impose these boundary conditions, we use the method of images. A singularity in the original square is reflected across each wall, and repeated reflections generate an infinite image lattice with period \(2L\) in both spatial directions. For a force \(\vF=(F_x,F_y)\) located at \(\mathbf{x}_s=(x_s, y_s)\), the image force density is
\begin{widetext}
\begin{equation}
\begin{aligned}
\vf(\vx)
=&\sum_{m,n=-\infty}^{\infty}(F_x,F_y)\delta(\vx-\mathbf{x}_s+2Ln\mathbf{e}_x+2Lm\mathbf{e}_y)\\
&+\sum_{m,n=-\infty}^{\infty}(-F_x,F_y)\delta(\vx-\mathbf{x}_s+2x_s\mathbf{e}_x+2Ln\mathbf{e}_x+2Lm\mathbf{e}_y)\\
&+\sum_{m,n=-\infty}^{\infty}(F_x,-F_y)\delta(\vx-\mathbf{x}_s+2y_s\mathbf{e}_y+2Ln\mathbf{e}_x+2Lm\mathbf{e}_y)\\
&+\sum_{m,n=-\infty}^{\infty}(-F_x,-F_y)\delta(\vx-\mathbf{x}_s+2x_s\mathbf{e}_x+2y_s\mathbf{e}_y+2Ln\mathbf{e}_x+2Lm\mathbf{e}_y).
\end{aligned}
\label{eq:image_force_density}
\end{equation}
\end{widetext}
Figure~\ref{fig:image_method} shows a schematic of the method of images used to satisfy the perfect-slip boundary conditions.

\subsection{Discrete Fourier representation of the confined Stokeslet}

We solve the image problem using a discrete Fourier transform on the doubled periodic domain \(2L\times2L\). Let \(N\) be the number of grid points along one side of the original square. One period of the image lattice contains
\begin{equation}
N_T=2(N-1)
\label{eq:grid_def}
\end{equation}
grid points in each direction, and the real-space grid spacing is
\begin{equation}
h=\frac{2L}{N_T}.
\label{eq:grid_spacing}
\end{equation}
The discrete position is
\begin{equation}
\vx=\left(\frac{2mL}{N_T},\frac{2nL}{N_T}\right),
\qquad
\mathbf{x}_s=\left(\frac{2m_s L}{N_T},\frac{2n_s L}{N_T}\right).
\end{equation}

In Fourier space, the Stokes equation is
\begin{equation}
|\vk|^2\mu\tilde{\vv}(\vk)+i\vk\tilde{p}(\vk)=\tilde{\vf}(\vk),
\label{eq:fourier_stokes}
\end{equation}
which gives the solenoidal projection
\begin{equation}
\tilde{\vv}(\vk)
=\frac{1}{\mu |\vk|^2}\tilde{\vf}(\vk)
\cdot
\left(\vI-\frac{\vk\vk}{|\vk|^2}\right),
\label{eq:fourier_velocity}
\end{equation}
where \(\vk=\frac{\pi}{L}(k,s)\). The zero mode \((k,s)=(0,0)\) is omitted.

The discrete force density over one period is
\begin{equation}
\begin{aligned}
\vf_{mn}
=&\frac{1}{h^2}\big[(F_x,F_y)\delta_{(m,n),(m_s,n_s)}+(-F_x,F_y)\delta_{(m,n),(-m_s,n_s)}\\
&+(F_x,-F_y)\delta_{(m,n),(m_s,-n_s)}+(-F_x,-F_y)\delta_{(m,n),(-m_s,-n_s)}\big].
\end{aligned}
\label{eq:discrete_force_density}
\end{equation}
The corresponding discrete Fourier coefficient is
\begin{equation}
\begin{aligned}
\tilde{\vf}_{ks}
=&\sum_{m,n=-N_T/2}^{N_T/2-1}\vf_{mn}
\ee^{-i\vk\cdot\vx}\\
=&\frac{1}{h^2}\bigg[(F_x,F_y)\ee^{-i\vk\cdot\vx_s}
+(-F_x,F_y)\ee^{-i\vk_2 \cdot\vx_s}\\
&+(F_x,-F_y)\ee^{-i\vk_3\cdot\vx_s}
+(-F_x,-F_y)\ee^{-i\vk_4\cdot\vx_s}\bigg].
\end{aligned}
\label{eq:DFTf}
\end{equation}
where
$\vk=\frac{\pi}{L}(k,s)$,
$\vk_2=\frac{\pi}{L}(-k,s)$,
$\vk_3=\frac{\pi}{L}(k,-s)$, and
$\vk_4=\frac{\pi}{L}(-k,-s)$.
Substitution into Eq.~\eqref{eq:fourier_velocity} gives
\begin{widetext}
\begin{equation}
\begin{aligned}
\tilde{\vv}_{ks}
=&\frac{1}{\mu |\vk|^2 h^2}
\bigg[(F_x,F_y)\ee^{-i\vk\cdot\vx_s}
+(-F_x,F_y)\ee^{-i\vk_2 \cdot\vx_s}\\
&+(F_x,-F_y)\ee^{-i\vk_3\cdot\vx_s}
+(-F_x,-F_y)\ee^{-i\vk_4\cdot\vx_s}\bigg]
\left(\vI-\frac{\vk\vk}{|\vk|^2}\right).
\end{aligned}
\label{eq:stokeslet_fourier_square}
\end{equation}
\end{widetext}
The confined Stokeslet velocity is obtained from the inverse discrete Fourier transform,
\begin{equation}
\vv_{mn}
=\frac{1}{N_T^2}
\sum_{k,s=-N_T/2}^{N_T/2-1}
\tilde{\vv}_{ks}\ee^{i\vk\cdot\vx}.
\label{eq:stokeslet_inverse_square}
\end{equation}

Equivalently, defining \(\vF=\alpha\ve=\alpha(e_1,e_2)\), the confined Green's function is
\begin{widetext}
\begin{equation}
\begin{aligned}
\mathbf{G}^{(c)}
=&\frac{4\pi\mu}{\alpha}\vv_{mn}\\
=&\frac{4\pi}{h^2 N_T^2}
\sum_{k,s=-N_T/2}^{N_T/2-1}
\frac{\ee^{i\vk\cdot\vx}}{|\vk|^2}\bigg[(e_1,e_2)\ee^{-i\vk\cdot\mathbf{x}_s}
+(-e_1,e_2)\ee^{-i\vk_2\cdot\mathbf{x}_s}
+(e_1,-e_2)\ee^{-i\vk_3\cdot\mathbf{x}_s}
+(-e_1,-e_2)\ee^{-i\vk_4\cdot\mathbf{x}_s}\bigg]
\left(\vI-\frac{\vk\vk}{|\vk|^2}\right),
\end{aligned}
\label{eq:greensquare}
\end{equation}
\end{widetext}
This confined Green's function is the starting point for constructing the stresslet and source-dipole contributions of the two-dimensional squirmer in a square domain.

\section{Regularized squirmer flow field in a square domain}
\label{app:squirmer}

\subsection{Relation to the two-mode squirmer model}

The two-dimensional squirmer is described by the tangential surface slip velocity
\begin{equation}
v^s_{\phi}(a,\phi)=\sum_{n=1}^{\infty}B_n\sin(n\phi).
\label{eq:surface_slip}
\end{equation}
In this work, only the first two modes are retained. The first mode is associated with a source-dipole contribution, while the second mode gives the stresslet contribution. In terms of the confined Green's function in Eq.~\eqref{eq:greensquare}, these components are
\begin{equation}
\vv_{B_1}^{(c)}(\mathbf{x})
=-\frac{a^2B_1}{2}\left(\frac{1}{2}\gradzero^2\mathbf{G}^{(c)}\right),
\label{eq:B1_source_operator}
\end{equation}
\begin{equation}
\vv_{B_2}^{(c)}(\mathbf{x})
=-B_2a\left(\ve\cdot\gradzero\mathbf{G}^{(c)}\right),
\label{eq:B2_stresslet_operator}
\end{equation}
where \(a\) is the swimmer radius. The swimmer type is characterized by
\begin{equation}
\beta=\frac{B_2}{B_1}.
\label{eq:beta_def}
\end{equation}
Thus \(|\beta|\to\infty\), \(\beta>0\), \(\beta<0\), and \(\beta=0\) correspond to a stresslet, puller, pusher, and source dipole, respectively.

\subsection{Confined stresslet}

Taking the derivative of Eq.~\eqref{eq:greensquare} with respect to \(\vxz\), the confined stresslet is
\begin{equation}
\begin{aligned}
\vv_{B_2}^{(c)}(\mathbf{x})=\vv^{B_2}_{mn}
&=-B_2a\left(\ve\cdot\gradzero\mathbf{G}^{(c)}\right)\\
&=\frac{1}{N_T^2}
\sum_{k,s=-N_T/2}^{N_T/2-1}
\ee^{i\vk\cdot\vx}\tilde{\vv}^{B_2}_{ks},
\end{aligned}
\label{eq:stresslet_square}
\end{equation}
where
\begin{widetext}
\begin{equation}
\begin{aligned}
\tilde{\vv}^{B_2}_{ks}
=&\frac{4\pi B_2ai}{|\vk|^2h^2}
\bigg[(e_1,e_2)(\vk\cdot\ve)\ee^{-i\vk\cdot\mathbf{x}_s}
+(-e_1,e_2)(\vk_2\cdot\ve)\ee^{-i\vk_2\cdot\mathbf{x}_s}\\
&+(e_1,-e_2)(\vk_3\cdot\ve)\ee^{-i\vk_3\cdot\mathbf{x}_s}
+(-e_1,-e_2)(\vk_4\cdot\ve)\ee^{-i\vk_4\cdot\mathbf{x}_s}\bigg]
\left(\vI-\frac{\vk\vk}{|\vk|^2}\right).
\end{aligned}
\label{eq:stresslet_fourier_square}
\end{equation}
\end{widetext}

\subsection{Confined source dipole}

The confined source dipole is
\begin{equation}
\begin{aligned}
\vv_{B_1}^{(c)}(\mathbf{x})
=\vv^{B_1}_{mn}
&=-\frac{a^2B_1}{2}\left(\frac{1}{2}\gradzero^2\mathbf{G}^{(c)}\right)\\
&=\frac{1}{N_T^2}
\sum_{k,s=-N_T/2}^{N_T/2-1}
\ee^{i\vk\cdot\vx}\tilde{\vv}^{B_1}_{ks},
\end{aligned}
\label{eq:source_square}
\end{equation}
where
\begin{equation}
\begin{aligned}
\tilde{\vv}^{B_1}_{ks}
=&\frac{\pi B_1a^2}{h^2}
\bigg[(e_1,e_2)\ee^{-i\vk\cdot\mathbf{x}_s}
+(-e_1,e_2)\ee^{-i\vk_2\cdot\mathbf{x}_s}\\
&+(e_1,-e_2)\ee^{-i\vk_3\cdot\mathbf{x}_s}
+(-e_1,-e_2)\ee^{-i\vk_4\cdot\mathbf{x}_s}\bigg]
\left(\vI-\frac{\vk\vk}{|\vk|^2}\right).
\end{aligned}
\label{eq:source_fourier_square}
\end{equation}

\subsection{Regularized confined squirmer flow}

The singularity at the swimmer position and the slow decay of the Fourier coefficients can lead to numerical instabilities when solving the Fokker--Planck equation. We therefore introduce exponential spectral regularization,
\begin{equation}
\tilde{\vv}_{ks}
\rightarrow
\tilde{\vv}_{ks}
\ee^{-\varepsilon |\mathbf{k}|/\pi},
\label{eq:regularization_factor}
\end{equation}
where \(\varepsilon\) is a short-distance regularization length.

The regularized stresslet and source-dipole fields are
\begin{equation}
\mathbf{\fvel}^{(r)}_{B_2}(\mathbf{x})=\vv^{B_2,(r)}_{mn}
=\frac{1}{N_T^2}\sum_{k,s=-N_T/2}^{N_T/2-1}
\ee^{i\vk\cdot\vx}
\tilde{\vv}^{B_2}_{ks}
\ee^{-\varepsilon |\mathbf{k}|/\pi},
\label{eq:regularized_stresslet_square}
\end{equation}
\begin{equation}
\mathbf{\fvel}^{(r)}_{B_1}(\mathbf{x})=\vv^{B_1,(r)}_{mn}
=\frac{1}{N_T^2}\sum_{k,s=-N_T/2}^{N_T/2-1}
\ee^{i\vk\cdot\vx}
\tilde{\vv}^{B_1}_{ks}
\ee^{-\varepsilon |\mathbf{k}|/\pi}.
\label{eq:regularized_source_square}
\end{equation}
The confined regularized squirmer flow field is therefore
\begin{equation}
\mathbf{\fvel}(\mathbf{x})=\mathbf{\fvel}^{(r)}_{B_1}(\mathbf{x})+\mathbf{\fvel}^{(r)}_{B_2}(\mathbf{x}).
\label{eq:squirmer_confined}
\end{equation}
In the simulations, we use
\begin{equation}
\varepsilon=\frac{3}{32}L,
\qquad
 a=\frac{\varepsilon}{2}.
\label{eq:epsilon_choice}
\end{equation}
Thus \(\varepsilon\) is interpreted as an effective swimmer diameter.

\subsection{Swimmer trajectories without hydrodynamic interactions}

\begin{figure}[b]
\centering
\includegraphics[width=0.65\columnwidth]{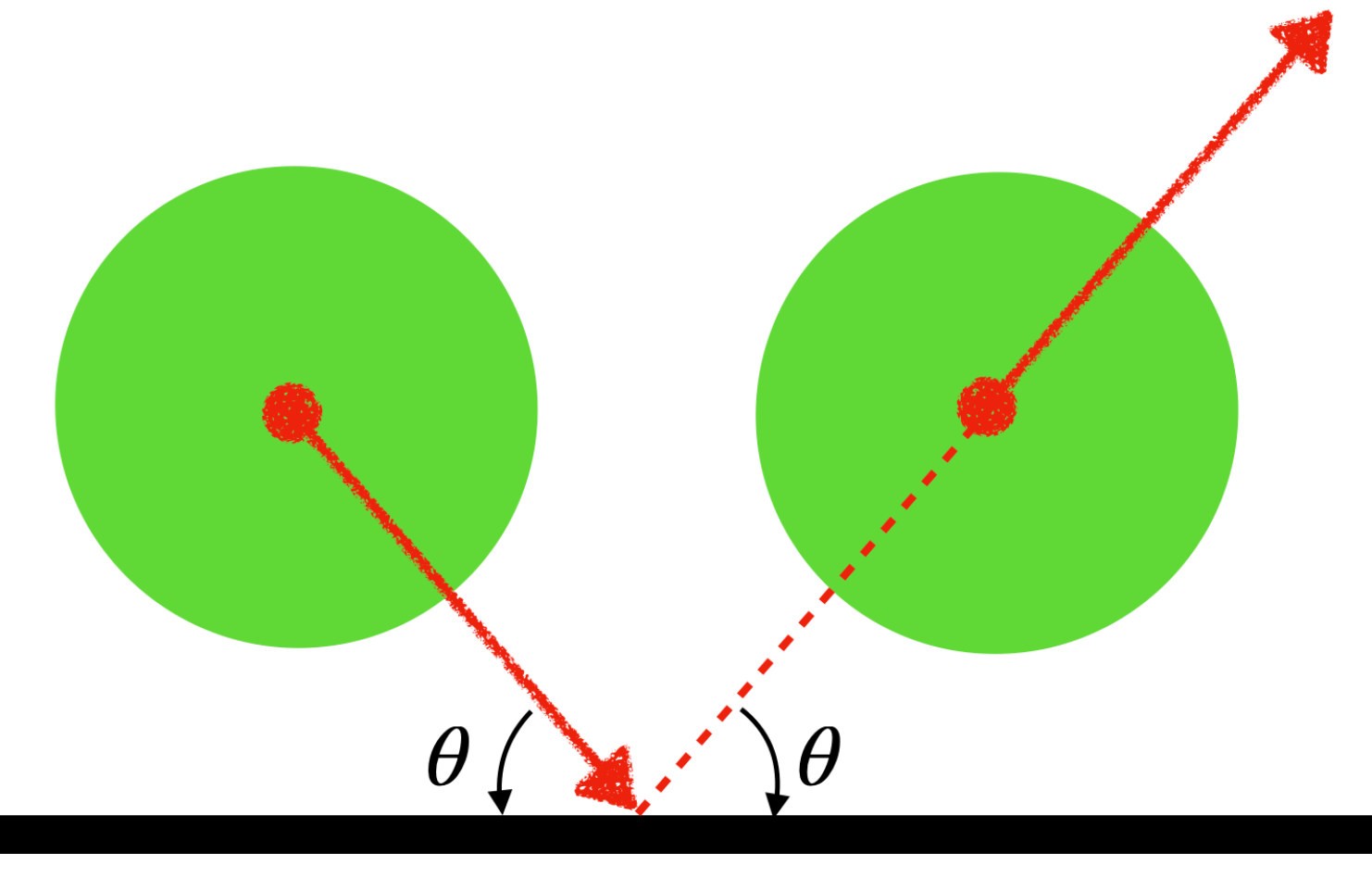}
\caption{Reflection rule for a swimmer reaching the wall. The translational velocity is $\mathbf{V}=B_1\mathbf{e}/2$, and the direction $\mathbf{e}$ is reflected at the boundary.}
\label{fig:reflection_rule}
\end{figure}

In the absence of hydrodynamic interactions, swimmers translate with the free-space squirmer velocity as shown in Eq.~\eqref{translational}.
The swimmer moves in a straight line until it reaches a wall, where its swimming direction is reflected geometrically: the normal component of \(\ve\) changes sign, while the tangential component remains unchanged. The schematic is shown in Fig.~\ref{fig:reflection_rule}.

\section{Hydrodynamic interactions between swimmers}
\label{app:hi}

In the main text, hydrodynamic interactions are included only where explicitly stated. In those simulations, each swimmer moves in the flow generated by the other swimmers. 
When hydrodynamic interactions are included, the rigid-body motion of each circular swimmer follows from Fax\'{e}n's laws,
\begin{align}
\mathbf{U}_i &= \mathbf{U}_s + \mathbf{u}_{\infty}(\mathbf{x}_i),\\
\boldsymbol{\Omega}_i &= \boldsymbol{\Omega}_s
   + \tfrac{1}{2}\,\nabla\times\mathbf{u}_{\infty}(\mathbf{x}_i),
\end{align}
where $\mathbf{u}_{\infty}(\mathbf{x}_i)$ is the flow generated by all other swimmers, evaluated at the position $\mathbf{x}_i$ of swimmer $i$ (its own field excluded). The free-space contributions are
$\mathbf{U}_s=\tfrac{B_1}{2}\mathbf{e}$ [Eq.~\eqref{translational}] and
$\boldsymbol{\Omega}_s=0$ between wall reflections; at a wall the orientation $\mathbf{e}$ is reflected geometrically, and $\boldsymbol{\Omega}_s$ accounts for this reorientation. The higher-order Fax\'{e}n corrections and wall-induced hydrodynamic corrections are neglected.

\bibliography{mixing.bib}% common bib file

\end{document}